\documentclass[aps,prl,twocolumn,groupedaddress,floatfix]{revtex4}

\usepackage{graphicx}
\usepackage{subfigure}
\usepackage{epsfig}
\usepackage{amsmath}
\usepackage{bbm}

\def\be{\begin{equation}}
\def\ee{\end{equation}}
\def\bea{\begin{eqnarray}}
\def\eea{\end{eqnarray}}

\begin{document}


\title{Cooper-pair propagation and superconducting correlations
in graphene}


\author{J. Gonz{\'a}lez$^1$ and E. Perfetto$^{2}$}
\affiliation{$^{1}$Instituto de Estructura de la Materia.
        Consejo Superior de Investigaciones Cient{\'\i}ficas.
        Serrano 123, 28006 Madrid. Spain.\\
             $^{2}$Consorzio Nazionale Interuniversitario per le Scienze
Fisiche della Materia, Universit\`a di Roma Tor Vergata, Via della
Ricerca Scientifica 1, 00133 Roma, Italy.}

\date{\today}

\begin{abstract}

We investigate the Cooper-pair propagation and the proximity
effect in graphene under conditions in which the distance $L$
between superconducting electrodes is much larger than the width
$W$ of the contacts. In the case of undoped graphene,
supercurrents may exist with a spatial decay proportional to
$W^2/L^3$. This changes upon doping into a $1/L^2$ behavior,
opening the possibility to observe a supercurrent over length
scales above $1 \; \mu$m at suitable doping levels. We also show
that there is in general a crossover temperature $T^* \sim v_F/k_B L$
that marks the onset of the strong decay of the supercurrent, 
and that corresponds to the scale below which the Cooper pairs are not 
disrupted by thermal effects during their propagation.

\end{abstract}

\maketitle

The recent interest in the physics of graphene has arisen from the
observation of a number of novel electronic
properties\cite{geim,kim}, which are the consequence of the
relativistic-like character of the electron quasiparticles. This
is certainly the case of the anomalous quantization of the Hall
conductivity\cite{ando,paco,gus}, as well as of the existence of a
finite lower bound in the conductivity at the charge neutrality
point\cite{paco,kat,mac,efe}. The graphene system is unique in
that the low-energy excitations have conical dispersion around
discrete Fermi points, being therefore governed by a Dirac
equation for massless chiral particles. The appearance of an
additional pseudo-spin quantum number intrinsic to the Dirac
spectrum has led to propose other unconventional effects, like the
selective transmission of electrons through $n$-$p$
junctions\cite{falk} or a peculiar form of Andreev reflection at
the metal-superconductor interface in graphene\cite{been}.

Recently the proximity effect has been investigated in graphene in two
different experiments\cite{delft,orsay}. In both of them a 2D carbon layer
has been contacted with superconducting electrodes, though the size and
aspect ratio of the respective samples have been quite different in
the two cases. Thus, while the separation of the electrodes in the
experiment of Ref. \onlinecite{delft} appears to be of the order of
$\sim 0.5 \; \mu$m, the distance between them in the report of Ref.
\onlinecite{orsay} seems to have a minimum value of about
$2.5 \; \mu$m. This may explain why a supercurrent has been measured
between the electrodes in the first case, while the most relevant
observation in the second experiment has been an abrupt drop in the
resistance, below the critical temperature of the superconducting
electrodes.

This crucial dependence on the distance between the
electrodes may be however somewhat surprising, taking into account
that supercurrents have been measured in carbon nanotubes placed
between superconducting contacts, with nanotube lengths as large as
$\approx 1.7 \; \mu$m \cite{kas}. It becomes then pertinent to analyze
the proximity effect in graphene, specially for large separation between
the superconducting electrodes, in order to find out the features
related to the 2D character of the material.
The Josephson effect has been studied before in graphene strips
with length $L$ small relative to their width and the superconducting
coherence length\cite{tit}. Here we will face the opposite situation,
in which the relevant signatures are dictated by the propagation of the
Cooper pairs in graphene and its dependence on thermal and interaction 
effects.

We investigate then the proximity effect in graphene, under conditions
in which the distance $L$ between the superconducting electrodes producing
the Cooper pairs is much
larger than the width $W$ of the contacts. More specifically, we will
consider that the hamiltonian for graphene with superconducting contacts 
along the coordinates $x_1 = 0$ and $x_2 = L$ is given by
\begin{eqnarray}
H  & = &  v_F \int d^2 r \; \Psi^{(a) \dagger}_{\sigma} (\mathbf{r})
   \mbox{\boldmath $\sigma$}^{(a)}\cdot\mbox{\boldmath $\partial$} \:
              \Psi^{(a)}_{\sigma} (\mathbf{r})
                                                  \nonumber      \\
  &  &  +   \sum_{j=1,2} t  \int_0^W dy \;
        \Psi^{(a) \dagger}_{\sigma} (x_j,y)  \Psi_{Sj,\sigma} (x_j,y)
                           +  {\rm h.c.}
\label{ham}
\end{eqnarray}
$\mbox{\boldmath $\sigma$}^{(a)}$ being different sets of Pauli
matrices for $a = 1, 2$ \cite{ando}. In the above expression,
a sum is implicit over the spin index $\sigma $,
as well as over the index $a$ accounting for the two
different valleys and corresponding Dirac spinors $\Psi^{(a)} $ at
opposite $K$ points in graphene. We will consider that the coupling to the
electron fields $\Psi_{S1}$ and $\Psi_{S2}$ in the superconducting electrodes
takes place along a segment of width $W$ in each case. For our purposes,
it will be enough to describe the superconductors in terms of the
order parameter $\Delta$ and the normal density of states $N$.

We are going to deal in particular with the case in which the time
of propagation of the Cooper pairs
between the contacts is much larger than $1/|\Delta |$.
This implies, equivalently, that the distance $L$ has to be much larger
than the superconducting coherence length $\xi $.
Then, a supercurrent may arise mainly from processes in which the Cooper
pairs tunnel from one of the superconductors to the 2D layer, propagating
to the other superconducting contact. The Josephson current $I_s$ can be
computed as the derivative of the free energy ${\cal F}$ with respect to the
difference $\chi $ between the respective phases $\chi_1 $ and $\chi_2 $
of the order parameters
in the two superconductors, $I_s = -2e \partial {\cal F}/\partial \chi$.
Under the assumption of a relative large $|\Delta |$, we can make the 
approximation
\begin{equation}
 \langle \Psi_{Sj,\sigma} (x_j,y;-i\tau_1) 
     \Psi_{Sj,-\sigma} (x_j,y;-i\tau_2)  \rangle  \approx 
     e^{i \chi_j} N \delta (\tau_1 - \tau_2 )
\end{equation}
where the statistical averages, at temperature $T$, are taken over ordered
products with respect to imaginary time $\tau $.
Then, the maximum supercurrent (critical current) becomes to lowest order 
in the tunneling amplitude
\begin{eqnarray}
I_c (T) & \approx &  2e N^2 t^4 W^2 \int_0^W dy_1 \int_0^W dy_2
     \int_0^{1/k_B T}  d \tau                         \nonumber     \\
 & &  \langle \Psi^{(a) \dagger}_{\uparrow} (0,y_1;0)
            \Psi^{(-a) \dagger}_{\downarrow} (0,y_1;0)   \nonumber     \\
  & &    \times     \Psi^{(a)}_{\uparrow} (L,y_2; -i\tau )
             \Psi^{(-a)}_{\downarrow} (L,y_2; -i\tau )   \rangle
\label{pert}
\end{eqnarray}
In deriving Eq. (\ref{pert}) we have already assumed that $L \gg W$, 
averaging over the separation of the electrons in the pair tunneling 
along the superconducting contact.

The determination of the critical current is therefore reduced to the
evaluation of the propagator of the Cooper pairs in the graphene layer.
The Fourier transform of this object, that we will denote by
$D(\mathbf{k}, i\overline{\omega})$, can be obtained from the standard
diagrammatics for Dirac fermions. Here we stress that, for the sake of
preserving the relativistic-like invariance, it is convenient to regularize
the diagrams at high energies by using a method that maintains the
space-time symmetry of the theory, like the analytic continuation in
the number of dimensions\cite{np}.
In particular, the
Cooper-pair propagator can be obtained in the noninteracting theory from
the convolution of two Dirac propagators, and at $T = 0$ it turns
out to be
\begin{equation}
 \left. D^{(0)} (\mathbf{k}, i \overline{\omega}) \right|_{T = 0} =
    -\frac{1}{8 v_F^2} \sqrt{v_F^2 \mathbf{k}^2 + \overline{\omega}^2}
\label{prop0}
\end{equation}
The temperature dependence of this object can be obtained using the
Matsubara formalism. In the static limit we get
\begin{eqnarray}
D^{(0)} (\mathbf{k}, 0)  & = &  - \frac{1}{2\pi v_F} |\mathbf{k}|
  \int_0^1 dx \sqrt{x(1-x)}                           \nonumber     \\
   &  &   \times      \tanh (v_F |\mathbf{k}| \sqrt{x(1-x)}/2k_B T)
                                                      \nonumber      \\
 &  &  - \frac{1}{\pi v_F^2} k_B T
  \int_0^1 dx                                         \nonumber       \\
  &  &   \times    \ln [ 2 \cosh (v_F |\mathbf{k}| \sqrt{x(1-x)}/2k_B T) ]
\label{prop}
\end{eqnarray}

The knowledge of the Cooper-pair propagator at zero frequency is enough to
compute the critical current in Eq. (\ref{pert}). Interaction effects
can be incorporated by summing up the different perturbative orders obtained
by iteration of the scattering of the particles in the pair. If we take
an average of the potential $V$ between the particles, we have
that the Cooper-pair propagator can be represented
in the T-matrix approximation by
\begin{equation}
D (\mathbf{k}, \omega) \approx \frac{D^{(0)} (\mathbf{k}, \omega)}
                {1 + V D^{(0)} (\mathbf{k}, \omega)}
\label{tm}
\end{equation}
We remark anyhow that the interaction effects cannot affect significantly
the propagation of the Cooper pairs at low temperatures.
When there is exchange of the valleys in the scattering of the electrons in
the pair, the Coulomb potential $V_C$ becomes suppressed by a large
momentum-transfer $2k_F $, down to $V_C \sim e^2/2k_F $. At the small values
of $|\mathbf{k}|$ relevant for the long-distance regime, the denominator
in Eq. (\ref{tm}) remains then very close to 1. In the opposite case of
no valley exchange, we observe that at low momentum-transfer
$V_C D^{(0)}$ has a relative strength of the order of $\sim e^2 /v_F$.
At low energies, this effective coupling is strongly renormalized, down
to a value which may be about one order of magnitude below the nominal
coupling\cite{prbr}. We have checked that, in practice, this makes very small
the difference between computing the critical current (\ref{pert}) from
either $D (\mathbf{k}, 0)$ or $D^{(0)} (\mathbf{k}, 0)$, at the relevant
temperatures in the experiments (of the order of $\sim 1$ K).

It becomes convenient to factor out from (\ref{pert}) the relative 
tunnel conductances of the interfaces, which are given each by the constant 
quantity $N t^2 W/v_F$ \cite{foot}. We will deal then
with the intrinsic 2D dependence of the critical current, which becomes in
terms of the Fourier transform of the Cooper-pair propagator
\begin{eqnarray}
I^{(2D)}_c (T)   &  \approx  &   2e W^2 v_F^2                  \nonumber      \\
      &  &     \times  \int_0^{\infty }
  \frac{dk}{2\pi } \; |\mathbf{k}| \; J_0 (|\mathbf{k}| L) D(\mathbf{k}, 0)
      e^{- |\mathbf{k}| /k_c }
\label{2D}
\end{eqnarray}
where $k_c$ is a short-distance cutoff, that we take of the order of
$\sim 1 \; {\rm nm}^{-1}$.

\begin{figure}
\begin{center}
\mbox{\epsfxsize 6.0cm \epsfbox{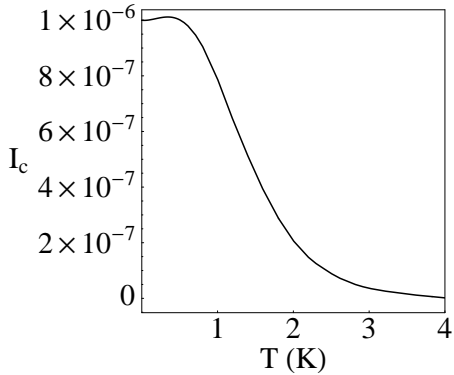}}
\end{center}
\caption{Plot of the critical current $I^{(2D)}_c (T)$ (in units of
$10^{-2} e v_F k_c$ $\approx 1.2 \; \mu$A) for $L = 5 \times 10^3 /k_c$
($\approx 2.5 \; \mu$m) and $W = 10^2 /k_c$ ($\approx 50$ nm).}
\label{one}
\end{figure}

The critical current computed from (\ref{2D}) displays a number of 
peculiar features. From the scaling of $D^{(0)} (\mathbf{k}, 0)$ in 
the zero-temperature limit, it is easily seen that $I^{(2D)}_c (0)$ must 
decay as $\sim v_F W^2 /L^3$ at large separation between the contacts. 
Another important property is that there is always a crossover temperature 
$T^* \sim v_F /k_B L $ which marks the onset of a power-law decay of the 
supercurrent at high temperatures. This is shown in Fig. \ref{one}, which 
represents the behavior of the critical current as a function of $T$ for 
$L = 5 \times 10^3 /k_c$ and $W = 10^2 /k_c$ . The dependence of 
$I^{(2D)}_c (T)$ is qualitatively similar to what is found in carbon 
nanotubes with long separation between superconducting contacts\cite{prl,jp}. 
The crossover temperature corresponds to the scale at which the Cooper pairs 
fail to propagate efficiently between the electrodes, as they become disrupted 
along the way by thermal effects.

The novel behavior of the supercurrents in graphene with respect to that
in the carbon nanotubes comes from the different scaling with length $L$
at low temperatures. This is given by a $1/L^3$ power-law behavior in
graphene, instead of the $1/L$ scaling of a noninteracting 1D
system\cite{prl}. We may consider for instance the sample described in Ref.
\onlinecite{orsay}, with an approximate separation between the electrodes
$L \approx 2.5 \; \mu$m. The plot in Fig. \ref{one} gives the theoretical
values of the critical current, in units of
$10^{-2} e v_F k_c \approx 1.2 \; \mu$A.
We observe that, in the low-temperature regime to the left of the crossover,
the expected critical currents should have a magnitude
$I^{(2D)}_c \sim 10^{-3}$ nA. Such an small scale may explain the difficulty 
to establish a supercurrent in graphene when there is a separation of the 
order of microns between the superconducting electrodes.

On the other hand, the crossover shown in Fig. \ref{one} has a remarkable
correspondence with the abrupt drop in the resistance reported in Ref.
\onlinecite{orsay}. This feature has been observed at a temperature of
about 1 K, and it does not seem to bear a direct relation to the critical
temperature $T_c$ ($ \approx 4$ K) of the superconducting electrodes.
We observe that the crossover displayed in Fig. \ref{one} corresponds to
a temperature of about 1.5 K, in fair agreement with the
position of the drop measured experimentally in the resistance. It is
therefore quite likely that this feature may be an indirect signature of
the temperature up to which the Cooper pairs are able to propagate without
disruption, along the $2.5 \; \mu$m-long path of the sample reported
in Ref. \onlinecite{orsay}.

The fast decay of the supercurrent computed from (\ref{2D}) can be
traced back to the vanishing density of states of graphene at the charge
neutrality point. Actually, the form of the propagator in (\ref{prop0})
is a direct consequence of the conical dispersion around the Fermi points
of graphene. One can therefore expect important changes in the Cooper-pair
propagation upon doping the electron system. This can be investigated
formally by introducing a finite chemical potential $\mu $, with the aim
of shifting the Fermi level away from the charge neutrality point.
At $\mu > 0$, the Dirac propagator can be written as\cite{chin}
\begin{eqnarray}
G^{(0)} (\mathbf{k}, \omega)  & =  &
  (  \omega + v_F \mbox{\boldmath $\sigma$} \cdot \mathbf{k} )
    [ \frac{-1}{-\omega^2 + v_F^2 \mathbf{k}^2 - i \epsilon}    \nonumber \\
  &  & + i \pi
       \frac{\delta (\omega - v_F |\mathbf{k}|)}{v_F |\mathbf{k}|}
         \theta (\mu - v_F |\mathbf{k}|)  ]
\label{diracmu}
\end{eqnarray}
which is a convenient alternative form of expressing the electron 
propagator
\begin{equation}
G^{(0)} (\mathbf{k}, \omega) = \frac{\theta ( \varepsilon(\mathbf{k}) - \mu )}
    {\omega - \varepsilon(\mathbf{k}) + i \epsilon}   
       +  \frac{\theta ( \mu - \varepsilon(\mathbf{k}) )}
    {\omega - \varepsilon(\mathbf{k}) - i \epsilon} 
\end{equation}
for a dispersion with two branches 
$\varepsilon(\mathbf{k}) = \pm v_F |\mathbf{k}|$.
The Cooper-pair propagator can be computed now from the convolution of two
Dirac propagators like (\ref{diracmu}). The result at $\omega = 0$ is
\begin{eqnarray}
D^{(0)} (\mathbf{k}, 0)  & = &  - \frac{1}{2\pi v_F^2} \mu  \;\;\;\;\;
  \;\;\;  \;\;\; \;\;\;  \;\;\;
           {\rm if}  \;\;\;   |\mathbf{k}| < 2 \mu    \nonumber    \\
        & = &  - \frac{1}{8 v_F} |\mathbf{k}|
    + \frac{1}{4\pi v_F} |\mathbf{k}|
        \arcsin \left( \frac{2\mu }{|\mathbf{k}|} \right)   \nonumber  \\
     &  &        - \frac{1}{2\pi v_F^2} \mu  \;\;\;\;\;
                {\rm if}  \;\;\;   |\mathbf{k}| > 2 \mu
\label{propmu}
\end{eqnarray}

\begin{figure}
\begin{center}
\mbox{\epsfxsize 6.0cm \epsfbox{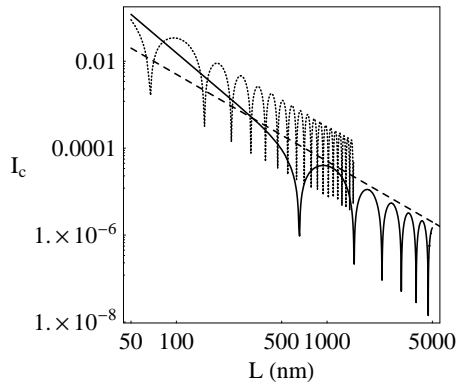}}
\end{center}
\caption{Plot of the zero-temperature critical current $I^{(2D)}_c (0)$ 
(in units of $10^{-2} e v_F k_c \approx 1.2 \; \mu$A) as a function of the 
distance $L$ (keeping $W = 10^2 /k_c \approx 50$ nm), for chemical potential 
$\mu = 1$ meV (full line) and $\mu = 0.01$ eV (dotted line). The dashed line
corresponds to a $1/L^2$ decay and it is drawn here as a reference of that
power-law behavior.}
\label{two}
\end{figure}

The slight change produced by the chemical potential in the infrared
behavior of $D^{(0)} (\mathbf{k}, 0)$ is enough to modify the long-distance
decay of the supercurrent. We have represented in Fig. \ref{two} the
result of evaluating $I^{(2D)}_c (0)$ from Eq. (\ref{2D}) with the Cooper-pair
propagator in (\ref{propmu}). It can be seen that, at $L \sim v_F/\mu$,
the power-law decay of the critical current changes from $1/L^3$ to
$1/L^2$. The supercurrents cannot be enhanced anyhow up to the magnitudes
that they reach in a 1D electron system, where the decay is given by a
$1/L$ dependence in the noninteracting theory. However, it may be worthwhile
to explore experimentally the consequences of doping the graphene layer,
up to levels where it can be affordable to measure supercurrents of the
order of $\sim 1$ nA for suitably small values of $L$, according to the
plot shown in Fig. \ref{two}.

We also remark that doping the graphene layer may be the way to obtain
experimental signatures of dynamical superconducting correlations. The
pertinent approach to address this question is to deal again with the
theory at nonvanishing chemical potential, starting from the Dirac
propagator in Eq. (\ref{diracmu}). Computing now the Cooper-pair propagator
at real frequency $\omega $, the possible superconducting instabilities
have to reflect as singularities in $D (0,\omega )$, as it happens with
the formation of bound electron pairs in the BCS theory\cite{agd}.

The evaluation of the Cooper-pair propagator at chemical potential
$\mu > 0$ gives the result
\begin{eqnarray}
D^{(0)} (0, \omega) & = &    -\frac{1}{4 \pi v_F^2} \omega
  \log \left| \frac{2\mu - \omega }{\omega } \right|      \nonumber  \\
    &  &  - \frac{1}{2 \pi v_F^2} \mu  - i \frac{1}{8 v_F^2} \omega
\label{dprop}
\end{eqnarray}
We may sum up again multiple scattering processes of the Cooper pairs
in the framework of the T-matrix approximation. We observe then that
$D(0, \omega )$ has a pole under conditions where the dominant
interaction is attractive, such that $V < 0$. The denominator 
$1 + V D^{(0)} (0, \omega )$ in (\ref{tm}) vanishes at values of 
the frequency $\omega $ given by the equation
\begin{equation}
|2\mu - \omega | = \omega
  \exp \left( \frac{4\pi v_F^2 - 2\mu V}{\omega V} \right)
\label{self}
\end{equation}
There is always a solution of Eq. (\ref{self}) for $\omega $ below and
close to $2 \mu $ when $V < 0$. Such a frequency marks the formation of 
bound electron pairs, that may take place for arbitrarily small values 
of $\mu $, despite the fact that the density of states of graphene
vanishes at $\omega = 0$. The main difference
with a conventional Fermi liquid is that the energy scale in front of
the exponential in Eq. (\ref{self}) may be rather small, of the order
of $\sim 2 \mu $.

In conclusion, we have studied the proximity effect in graphene when the
distance between the superconducting contacts is much larger than their 
width. 
We have seen that, in the case of undoped graphene, the
supercurrents have a fast spatial decay, proportional to $W^2/L^3$.
This strong dependence on $L$ reminds of the behavior of the critical
current in long diffusive junctions\cite{dubos}. On the other hand, 
it is unlikely that the experimental measures may be strongly affected
by disorder in a graphene layer, since backscattering arises 
only from small-range scatterers with a size not larger than the
lattice constant\cite{ando2}. We have actually seen that the results 
reported in Ref. \onlinecite{orsay} are consistent with a crossover
to the strong decay of the critical current at a temperature
$T^* \sim v_F/k_B L$, and not at the much smaller scale given in a 
diffusive junction in terms of the diffusion constant ${\cal D}$ by 
$T^* \sim {\cal D}/k_B L^2$.

We have seen that shifting the Fermi level away from the charge 
neutrality point changes the dependence of the critical current into 
a $1/L^2$ behavior, opening the possibility to observe a supercurrent 
over length scales above $1 \; \mu$m at suitable doping levels. 
We have shown that this becomes feasible anyhow below
the crossover temperature $T^*$ marking the onset of the power-law decay 
of the supercurrent, and that corresponds to the scale above which the 
Cooper pairs are increasingly disrupted by thermal effects during their
propagation.

We thank H. Bouchiat and F. Guinea for useful comments on the paper.
The financial support of the Ministerio de Educaci\'on y Ciencia
(Spain) through grant FIS2005-05478-C02-02 is gratefully
acknowledged. E.P. is also financially supported by CNISM (Italy).


\begin{thebibliography}{99}
\bibitem{foot}
It can be easily seen that the relative conductance 
$N t^2 W/v_F$ is a dimensionless quantity, as the electron fields 
$\Psi^{(a)}$ in graphene have the dimension of the inverse of a length 
according to Eq. (\ref{ham}).
\end{thebibliography}
\end{document}